# Bipartite entanglement in AJL's algorithm for three-strand braids

Ri Qu[1,2], Weiwei Dong[1,2], Juan Wang[1,2], Yanru Bao[1,2], Yin Song[1,2], Dawei Song[1,2,3]

*1 School of Computer Science and Technology, Tianjin University, Tianjin 300072, China*
*2 Tianjin Key Laboratory of Cognitive Computing and Application, Tianjin 300072, China*
*3 Faculty of Mathematics, Computing and Technology, The Open University, Milton Keynes MK7 6AA, United Kingdom*

Aharonov, Jones, and Landau [Algorithmica 55, 395 (2009)] have presented a polynomial quantum algorithm for approximating the Jones polynomial. We investigate the bipartite entanglement properties in AJL's algorithm for three-strand braids. We re-describe AJL's algorithm as an equivalent algorithm which involves three work qubits in some mixed state coupled to a single control qubit. Furthermore, we use the Peres entanglement criterion to study the entanglement features of the state before measurements present in the re-described algorithm for all bipartitions. We show that the state is a product state relative to the bipartition between the first work qubit and the others. And it has no entanglement between the control qubit and work ones. We also prove a sufficient and necessary condition for its entanglement between the second (third) work qubit and the others. Moreover, we discuss the relation between its bipartite entanglement and elementary crossings in the three-strand braid group. We find that braids whose trace closures are topologically identical may have different entanglement properties in AJL's algorithm.

I. Introduction

A celebrated result in quantum computation [1] is the discovery of some quantum algorithms [2-6] able to solve problems faster than any known classical algorithm. Although it is well known that *entanglement* [7] represents an essential ingredient in quantum communication, its role in the speed-up of quantum computation is not yet fully understood. In particular, it is of great interest to investigate the role of entanglement in quantum algorithms. In Shor's algorithm entanglement was proved to be necessary to achieve exponential speed-up with quantum resources [8]. Ref. [9] showed that multipartite entangled states are employed in the Deutsch-Jozsa algorithm and in the first step of Grover's algorithm. Moreover, more recently it was shown that genuine multipartite entanglement is always present in Grover's algorithm and the dynamics of entanglement shows the behavior of scale invariance [10].

Knot invariants help to solve a fundamental problem in knot theory: determining whether two links (including knots) are topologically different, up to ambient isotopy. Tow links are said to be *topologically identical* if one can be maneuvered into the other by a sequence of Reidemeister moves. A *knot invariant* is a function on links which has the same value for topologically identical links. In

1984, Jones [11] discovered a new knot invariant, called the *Jones polynomial*, which is a Laurent polynomial with integer coefficients. In addition to the important role it has played in low dimensional topology, the Jones polynomial has found applications in numerous fields, e.g., DNA recombination [12], statistical physics [13], etc. Unfortunately, exact evaluation of the Jones polynomial at all but a few points is hard for the complexity class $\#P$ [14]. AJL's algorithm [6] can approximate the Jones polynomial of an $n$-strand braid at any primitive root of unity on a quantum computer in polynomial time. In this paper, our main aim is to investigate in detail the bipartite entanglement properties in AJL's algorithm for three-strand braids.

This paper is organized as follows. In Sec. II we will review some notations and definitions, including the braid group, the Temperley-Lieb algebra, the Jones polynomial, the path model representation, etc. Moreover, we will describe AJL's algorithm and the Peres entanglement criterion. We will re-describe AJL's algorithm as a mix-state algorithm in Sec. III. Then we will prove that these tow algorithm are equivalent. In Sec. IV we will study the entanglement properties in the re-described algorithm by means of the Peres entanglement criterion. We will summarize our conclusions in Sec. V.

II. Preliminaries

*Links* (i.e. circles embedded in $\mathbb{R}^3$) can be described in the discrete language of braid groups. As shown in Fig. 1, a *braid* is a series of strands crossings over and under each other with loose ends at both the top and bottom. Every braid might be converted into a link by the *trace closure* which connects the top and bottom ends of the braid in sequential order.

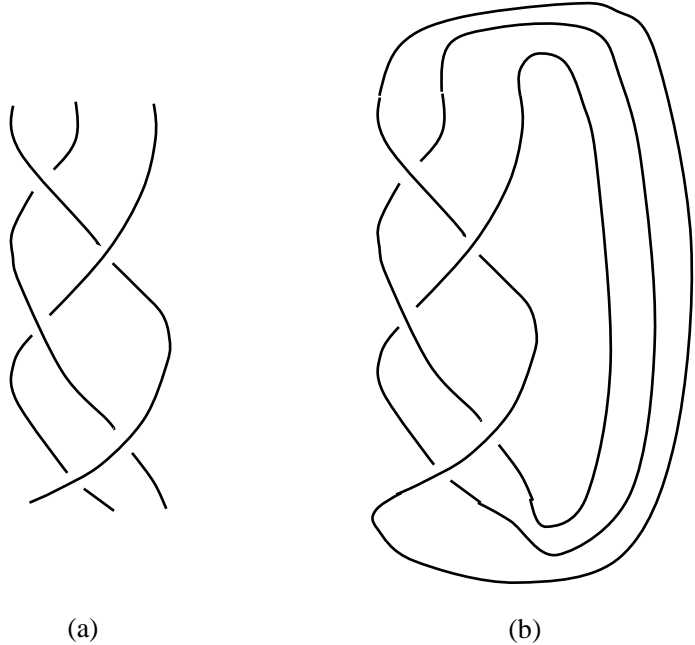

(a)          (b)

Fig. 1. A three-strand braid and its trace closure. For convenience, we neglect the direction of the link in (b) which is the trace closure of the braid in (a).

The *braid group* for three strands $B_3$ is generated by $\sigma_1$ and $\sigma_2$ in Fig. 2 that denote elementary crossings where $\sigma_j$ indicates the $j$th strand crossing over the $(j+1)$th strand and $\sigma_j^{-1}$ indicates the $j$th strand crossing under the $(j+1)$th strand. These two elementary crossings satisfy the following relation:

$$\sigma_1\sigma_2\sigma_1 = \sigma_2\sigma_1\sigma_2. \tag{1}$$

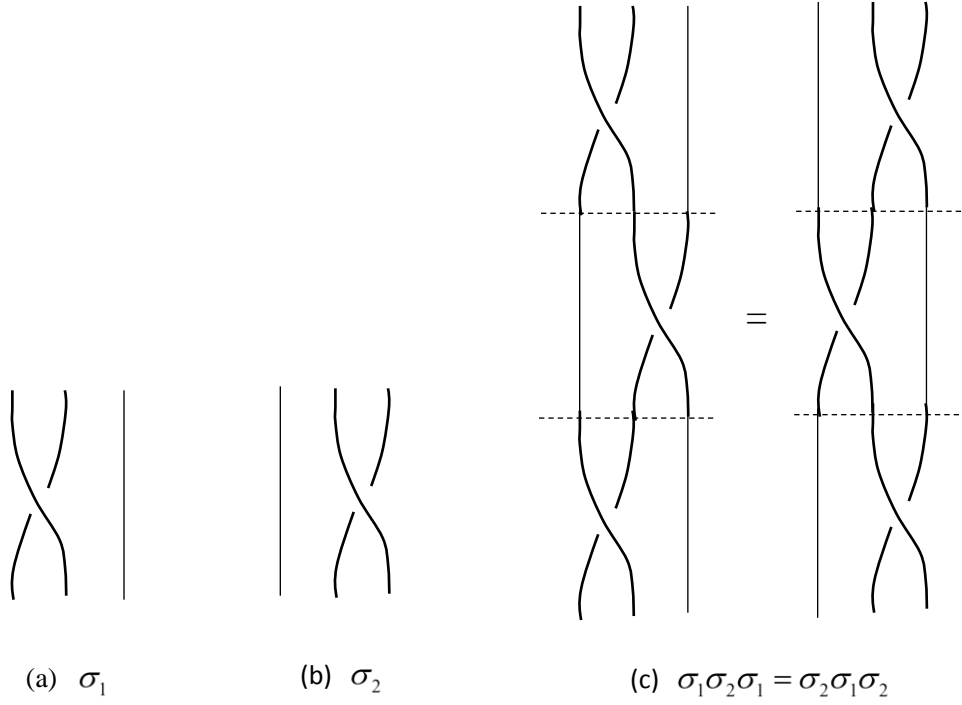

(a) $\sigma_1$     (b) $\sigma_2$     (c) $\sigma_1\sigma_2\sigma_1 = \sigma_2\sigma_1\sigma_2$

Fig. 2. Three-strand braid group and its elementary crossings.

Given a complex number $d$, define the *Temperley-Lieb algebra* $TL_3(d)$ to be the algebra generated by $\mathbf{1}$, $E_1$, $E_2$ with the relations:

$$E_1E_2E_1 = E_1, \quad E_2E_1E_2 = E_2, \text{ and } E_j^2 = dE_j \text{ for } j=1,2, \tag{2}$$

where $\mathbf{1}$ is the identity element. Define a representation $\tau_A$ of the braid group $B_3$ inside $TL_3(d)$ as follows. For $j=1,2$,

$$\tau_A(\sigma_j) = AE_j + A^{-1}\mathbf{1} \text{ and } \tau_A(\sigma_j^{-1}) = A^{-1}E_j + A\mathbf{1}, \tag{3}$$

where $d = -A^{-2} - A^2$. Moreover, there is a well-known geometric description of $TL_3(d)$ due to Kauffman [15] as shown in Fig. 3. Denote by $B^{tr}$ the trace closure of a three-strand braid $B$. Then the Jones polynomial $V_{B^{tr}}(t)$ of $B^{tr}$ at $t = A^{-4}$ is equal to

$$V_{B^{tr}}(t) = V_{B^{tr}}(A^{-4}) = (-A)^{-3w(B^{tr})} d^2 tr[\tau_A(B)] \tag{4}$$

where $w(L)$ is the writhe of the link $L$ and $tr(E)$ is the *Markov trace* of $E \in TL_3(d)$.

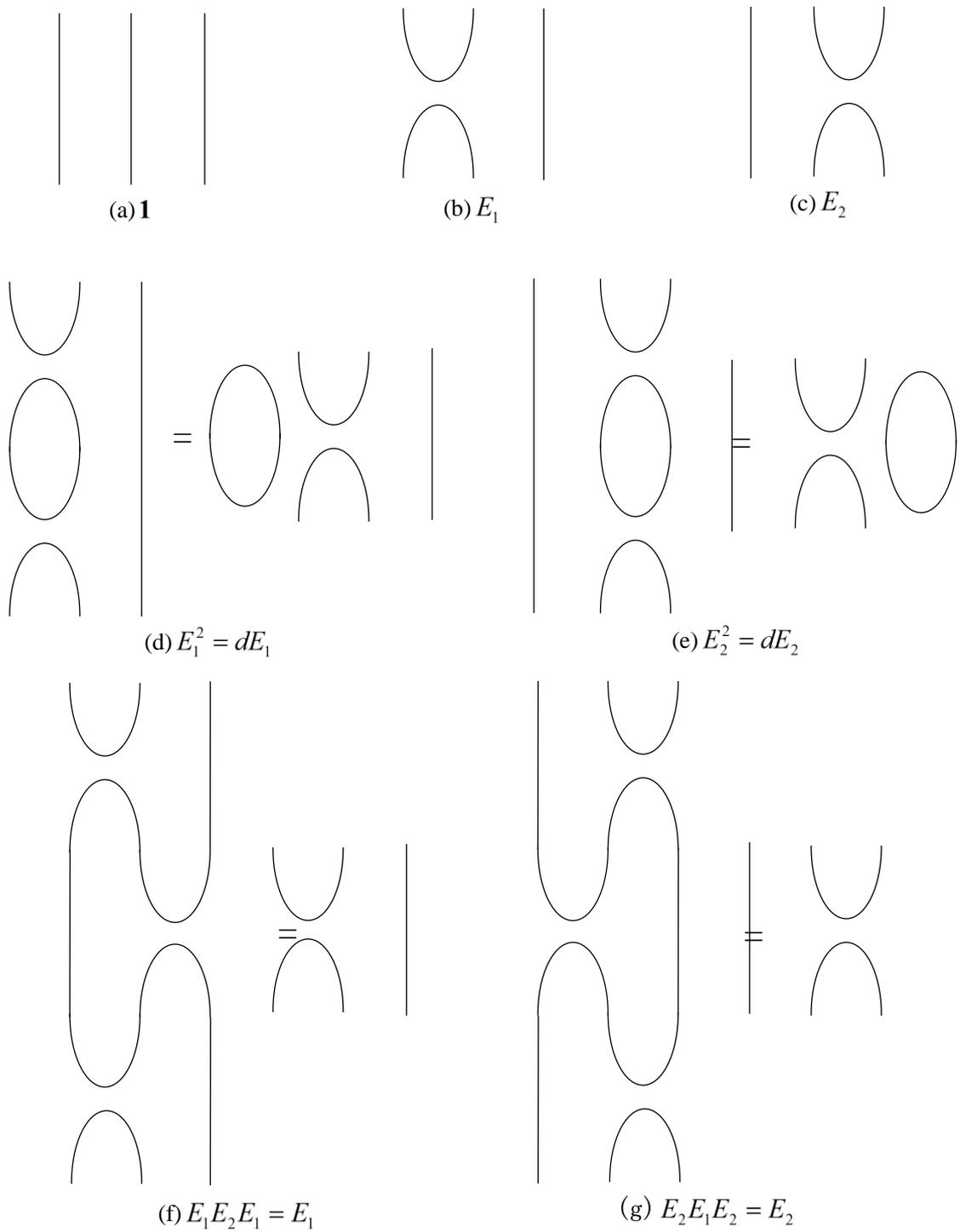

Fig. 3. The geometric description of the Temperley-Lieb algebra $TL_3(d)$.

For an integer $k \geq 5$, denote by $G_k$ a straight line graph with $(k-2)$ segments and $(k-1)$ vertices as shown in Fig. 4. Define $P_{3,k}$ to be the set of all paths of three steps on the graph $G_k$ beginning at the leftmost vertex. The path $p \equiv p_1 p_2 p_3$ is a three-bit string where $p_j$ indicates the direction of $j$-th step, that is, a 0 means taking one step to left and a 1 means taking one step to the right. Note that $p_1 = 1$ because all paths begin at leftmost vertex. It is clear that $P_{3,k} = \{101, 110, 111\}$. Let $p|_j$ denote the restriction of a path $p$ to its first $(j-1)$ bits, that is, $p|_j \equiv p_1 ... p_{j-1}$. Denote by $\ell(p)$ the vertex in $G_k$ which the path $p$ ends at. Let $\lambda_j = \sin(\pi j/k)$ and $d = 2\cos(\pi/k)$.

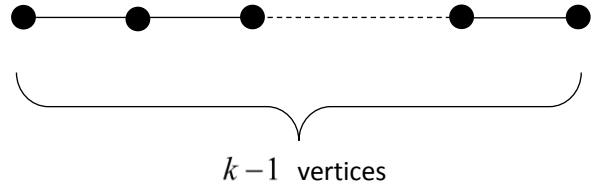

$k-1$ vertices

Fig. 4. A straight line graph $G_k$ with $(k-2)$ segments and $(k-1)$ vertices.

Then define a representation $\Phi$ from $TL_3(d)$ inside operators on three qubits as follows.

$$\Phi(E_1) = \frac{\lambda_2}{\lambda_1}|101\rangle\langle 101| = d|101\rangle\langle 101|, \qquad (5)$$

$$\Phi(E_2) = \frac{\lambda_1}{\lambda_2}|101\rangle\langle 101| + \frac{\sqrt{\lambda_1 \lambda_3}}{\lambda_2}|101\rangle\langle 110| + \frac{\sqrt{\lambda_1 \lambda_3}}{\lambda_2}|110\rangle\langle 101| + \frac{\lambda_3}{\lambda_2}|110\rangle\langle 110|$$

$$= \frac{1}{d}|101\rangle\langle 101| + \frac{\sqrt{d^2-1}}{d}|101\rangle\langle 110| + \frac{\sqrt{d^2-1}}{d}|110\rangle\langle 101| + \frac{d^2-1}{d}|110\rangle\langle 110| \quad (6)$$

Note that $\Phi(\mathbf{1}) = I^{\otimes 3}$ where $I$ is the identity operator on one qubit. It is clear that $\Phi(E_1)$ and $\Phi(E_2)$ are Hermitian. Thus it is known that if $A = ie^{-\pi i/2k}$ then $\Phi \circ \tau_A(\sigma_j)$ and $\Phi \circ \tau_A(\sigma_j^{-1})$ are unitary for $j = 1, 2$ [6]. Given a braid $B \in B_3$ with $m$ crossings, there are $\chi_{j_1}, \chi_{j_2}, ..., \chi_{j_m} \in \{\sigma_1, \sigma_2, \sigma_1^{-1}, \sigma_2^{-1}\}$ such that $B = \chi_{j_1} \chi_{j_2} ... \chi_{j_m}$. Let $U_B \equiv \Phi \circ \tau_A(B)$. Then It is clear that $U_B = U_{\chi_{j_1}} U_{\chi_{j_2}} ... U_{\chi_{j_m}}$. Ref. [6] shows that the Markov trace of $\tau_A(B)$ can be written into

$$tr[\tau_A(B)] = \sum_{p \in P_{3,k}} \Pr(p) \langle p | U_B | p \rangle$$

$$= \frac{1}{d^2}\langle 101|U_B|101\rangle + \frac{1}{d^2}\langle 110|U_B|110\rangle + \left(1-\frac{2}{d^2}\right)\langle 111|U_B|111\rangle \quad (7)$$

where $\Pr(p) = \lambda_{\ell(p)} \Big/ \sum_{q\in P_{3,k}} \lambda_{\ell(q)}$.

AJL's algorithm is used to approximate the Jones polynomial of a braid at any primitive root of unity. For a three-strand braid $B$ with $m$ crossings, this algorithm is described as follows.

i) Repeat for $j=1$ to $poly(m,k)$:

   a) Classically pick a random path $p \in P_{3,k}$ with probability $\Pr(p)$;

   b) Run the Hadamard test with measurement $X$ as shown in Fig. 5. Firstly, the control qubit $c$ and work qubits $\{1,2,3\}$ are respectively initialized to $|0\rangle$ and $|p\rangle$, that is, the initial state is $|0\rangle|p\rangle$. Apply the Hadamard gate $H$ to the control qubit to get the state $|+\rangle|p\rangle$ where $|+\rangle \equiv \frac{1}{\sqrt{2}}(|0\rangle+|1\rangle)$. Then apply the controlled-$U_B$ to create $\frac{1}{\sqrt{2}}(|0\rangle|p\rangle + |1\rangle \otimes U_B|p\rangle)$. Measure the control qubit using $X$ to get the measurement result $x_j$.

   c) Do the same as b) but for the measurement result $y_j$ by means of using measurement $Y$ as shown in Fig. 5.

ii) Let $r$ be the average over all $x_j + iy_j$. Output $(-A)^{-3w(B^{tr})} d^2 r$.

It is known that the expectation value of $x_j (y_j)$ in AJL's algorithm is the real (imaginary) part of $\langle p|U_B|p\rangle$. Combined with classical random sampling, the average over all $x_j + iy_j$ approximates the value of $\sum_{p\in P_{3,k}} \Pr(p)\langle p|U_B|p\rangle$.

A mixed state $\rho$ on a Hilbert space $H = H_M \otimes H_N$ is said to be *separable* with respect to the bipartition $\{M;N\}$ if and only if it can be decomposed as

$$\rho = \sum_j p_j |e_j\rangle\langle e_j| \otimes |f_j\rangle\langle f_j|, \quad \sum_j p_j = 1 \quad (8)$$

where $|e_j\rangle$ and $|f_j\rangle$ are respectively the normalized pure states on $H_M$ and $H_N$ [16]. If $\rho$ is not separable, it is *entangled*. If the partial transpose of $\rho$ denoted $\overline{\rho}^M$ has no negative eigenvalue, the state $\rho$ is said to be *positive partial transpose (PPT)*. Otherwise, we call $\rho$ a *non-PPT state*. A sufficient condition for entanglement, called the *Peres entanglement criterion*, is that $\overline{\rho}^M$ has a negative eigenvalue [17]. This means that non-PPT states are entangled, that is, separable states is PPT. If there exist two pure states $|\theta\rangle$, $|\varphi\rangle$ such that $\langle\theta|\overline{\rho}^M|\varphi\rangle \neq 0$ and

$\langle\varphi|\bar{\rho}^M|\varphi\rangle=0$, then $\rho$ is non-PPT [18]. Moreover, any PPT state $\rho$ is not *distillable* (namely, one cannot distill a singlet state out of many copies of $\rho$ by means of local operations and classical communication [19]). Clearly, entangled PPT states are not distillable. If $\rho$ is not distillable, then $ran(\rho) \geq \max\{ran[Tr_M(\rho)], ran[Tr_N(\rho)]\}$ where $ran(\rho)$ is the rank of $\rho$ and $Tr_M(\rho)$ is the reduced density operator on $H_M$ of $\rho$ [20]. Moreover, it is known that $\rho$ is separable if and only if it is PPT when it satisfies $ran[Tr_M(\rho)]ran[Tr_N(\rho)] \leq 6$ [21].

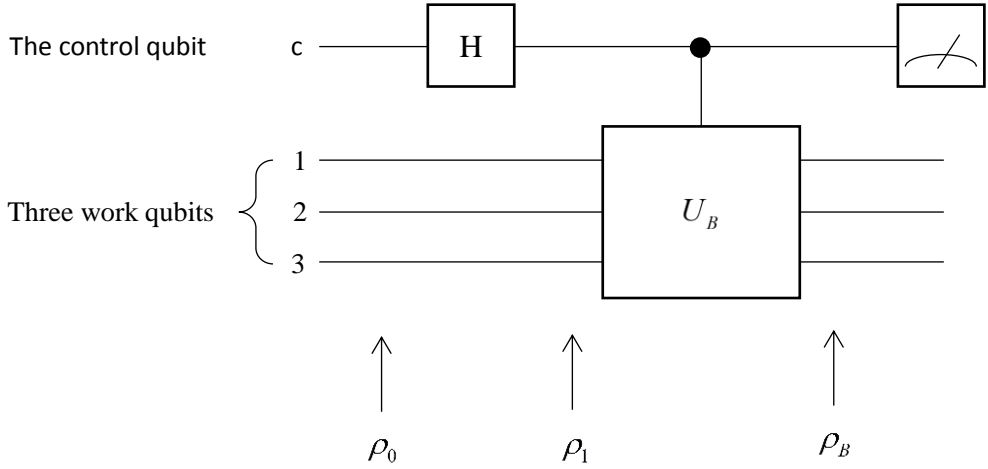

Fig. 5. The Hadamard test. In AJL's algorithm, the control qubit $c$ and work qubits $\{1,2,3\}$ are respectively initialized to $|0\rangle$ and $|p\rangle$, that is, the initial state is $|0\rangle|p\rangle$. In the re-described algorithm shown in Sec. III, the initial state is $\rho_0 = |0\rangle\langle 0| \otimes \phi$ where $\phi$ is defined in (9). Moreover, in DQC1 [22], the initial state is $|0\rangle\langle 0| \otimes I^{\otimes 3}$.

III. Re-described AJL's algorithm

It is clear that AJL's algorithm includes classical random samplings and the circuit based on pure states. In this section this algorithm are re-described as an equivalent algorithm only including the circuit based on mixed states. Let $B$ be a three-strand braid with $m$ crossings. Then the re-described algorithm is shown as follows.

   i') Repeat for $j=1$ to $poly(m,k)$:

a') Run the Hadamard test with measurement $X$ as shown in Fig. 5. Firstly, the control qubit $c$ and work qubits $\{1,2,3\}$ are respectively initialized to $|0\rangle\langle 0|$ and $\phi$, that is, the initial state is $\rho_0 = |0\rangle\langle 0| \otimes \phi$ where

$$\phi = \sum_{p \in P_{3,k}} \Pr(p)|p\rangle\langle p| = \frac{1}{d^2}|101\rangle\langle 101| + \frac{1}{d^2}|110\rangle\langle 110| + \left(1 - \frac{2}{d^2}\right)|111\rangle\langle 111|. \qquad (9)$$

Apply the Hadamard gate $H$ to the control qubit to get the state $\rho_1 = |+\rangle\langle +| \otimes \phi$. Then apply the controlled-$U_B$ to create the state

$$\rho_B = \frac{1}{2}\left(|0\rangle\langle 0| \otimes \phi + |0\rangle\langle 1| \otimes \phi U_B^\dagger + |1\rangle\langle 0| \otimes U_B \phi + |1\rangle\langle 1| \otimes U_B \phi U_B^\dagger\right) \qquad (10)$$

Measure the control qubit using $X$ to get the measurement result $x'_j$.

b') Do the same as a') but for the measurement result $y'_j$ by means of using measurement $Y$ as shown in Fig. 5.

ii') Let $r'$ be the average over all $x'_j + i y'_j$. Output $(-A)^{-3w(B^{tr})} d^2 r'$.

*Lemma 1.* The above algorithm is equivalent to AJL's algorithm.

Proof. We only need to prove that the expectation value of $x'_j (y'_j)$ is the real (imaginary) part of $\sum_{p \in P_{3,k}} \Pr(p)\langle p|U_B|p\rangle$. In fact, the expectation value of $x'_j$ is equal to

$$Tr\left[(X \otimes I^{\otimes 3})\rho_B\right] = Tr\left[\frac{1}{2}\left(|1\rangle\langle 0| \otimes \phi + |1\rangle\langle 1| \otimes \phi U_B^\dagger + |0\rangle\langle 0| \otimes U_B \phi + |0\rangle\langle 1| \otimes \phi\right)\right]$$

$$= Tr\left[\frac{1}{2}\left(|1\rangle\langle 1| \otimes \phi U_B^\dagger + |0\rangle\langle 0| \otimes U_B \phi\right)\right]$$

$$= \frac{1}{2}\left[Tr(\phi U_B^\dagger) + Tr(U_B \phi)\right]$$

$$= \frac{1}{2}\left(\sum_{p \in P_{3,k}} \Pr(p)\langle p|U_B^\dagger|p\rangle + \sum_{p \in P_{3,k}} \Pr(p)\langle p|U_B|p\rangle\right)$$

$$= \mathrm{Re}\left(\sum_{p \in P_{3,k}} \Pr(p)\langle p|U_B|p\rangle\right)$$

where $Tr(\rho)$ is the standard trace of $\rho$ and $\mathrm{Re}(r)$ is the real part of a complex number $r$. Similarly, we can prove that $Tr\left[(Y \otimes I^{\otimes 3})\rho_B\right]$ equals to the imaginary part of $\sum_{p \in P_{3,k}} \Pr(p)\langle p|U_B|p\rangle$. □

IV. Entanglement in the re-described algorithm

It is clear that $\rho_0$ and $\rho_1$ present in the above re-described algorithm are full separable. In this section, we use the Peres entanglement criterion to investigate the entanglement properties of the state $\rho_B$ in (10) for all bipartitions. Firstly, we discuss the entanglement features of $\rho_B$ with respect to the bipartition $\{\{1\};\{c,2,3\}\}$. According to (3), (5), and (6), the unitary operator $U_B$ in the algorithm can be written into the following form

$$U_B = \sum_{\substack{p,q \in P_{3,k} \\ \wedge d(p)=d(q)}} u_{B|p,q} |p\rangle\langle q|$$

$$= u_{B|101,101}|101\rangle\langle 101| + u_{B|110,110}|110\rangle\langle 110| + u_{B|111,111}|111\rangle\langle 111|$$

$$+ u_{B|101,110}|101\rangle\langle 110| + u_{B|110,101}|110\rangle\langle 101| \quad (11)$$

where $u_{B|p,q}$ are complex numbers and $d(p)$ is the Hamming weight of $p$, i.e., the number of 1s in $p$. By (9) and (11), it is true that $U_B \phi U_B^\dagger = \phi$. Thus the state $\rho_B$ in (10) can be written into

$$\rho_B = \frac{1}{2}\left(|0\rangle\langle 0|\otimes\phi + |0\rangle\langle 1|\otimes\phi U_B^\dagger + |1\rangle\langle 0|\otimes U_B\phi + |1\rangle\langle 1|\otimes\phi\right). \quad (12)$$

Then the following lemma can be obtained according to (11) and (12).

*Lemma 2.* The state $\rho_B$ can be written into a product state $|1\rangle_1\langle 1| \otimes Tr_{\{c,2,3\}}(\rho_B)$ where $Tr_{\{c,2,3\}}(\rho_B)$ is the reduced density operator on the qubits $\{c,2,3\}$ of $\rho_B$.

For the control qubit and three work qubits $\{1,2,3\}$, there are only four different bipartitions: $\{\{1\};\{c,2,3\}\}$, $\{\{c\};\{1,2,3\}\}$, $\{\{2\};\{c,1,3\}\}$, and $\{\{3\};\{c,1,2\}\}$. In fact, the bipartition $\{\{c,1\};\{2,3\}\}$ is the same as $\{\{c\};\{1,2,3\}\}$ for bipartite entanglement of $\rho_B$ according to the above lemma. Similarly, the bipartition $\{\{1,2\};\{c,3\}\}$ ( $\{\{1,3\};\{c,2\}\}$ ) is the same as $\{\{2\};\{c,1,3\}\}$ ( $\{\{3\};\{c,1,2\}\}$ ).

*Proposition 3.* The state $\rho_B$ is PPT with respect to the bipartition $\{\{c\};\{1,2,3\}\}$.

Proof. According to (12), the partial transpose $\overline{\rho}_B^{\{c\}}$ of the state $\rho_B$ can be written into

$$\overline{\rho}_B^{\{c\}} = \frac{1}{2}\left(|0\rangle\langle 0|\otimes\phi + |0\rangle\langle 1|\otimes U_B^*\phi + |1\rangle\langle 0|\otimes\phi U_B^T + |1\rangle\langle 1|\otimes\phi\right) \quad (13)$$

The characteristic equation of $\overline{\rho}_B^{\{c\}}$ can be obtained as follows.

$$\left|\lambda I^{\otimes 4}-\overline{\rho}_{B}^{\{c\}}\right|=\left|\lambda I^{\otimes 3}-\frac{1}{2}\phi\right|\left|\left(\lambda I^{\otimes 3}-\frac{1}{2}\phi\right)-\frac{1}{4}\phi U_{B}^{T}\left(\lambda I^{\otimes 3}-\frac{1}{2}\phi\right)^{-1}U_{B}^{*}\phi\right|$$

$$=\left|\lambda I^{\otimes 3}-\frac{1}{2}\phi\right|\left|\left(\lambda I^{\otimes 3}-\frac{1}{2}\phi\right)-\frac{1}{4}\phi\left(\lambda I^{\otimes 3}-\frac{1}{2}\phi\right)^{-1}\phi\right|$$

$$=\lambda^{13}\left(\lambda-\frac{1}{d^2}\right)\left(\lambda-\frac{1}{d^2}\right)\left(\lambda-1+\frac{2}{d^2}\right)=0.$$

From the above equation, it is clear that eigenvalues of $\overline{\rho}_{B}^{\{c\}}$ are the same as ones of $\rho_B$. Thus The state $\rho_B$ is PPT with respect to the bipartition $\{\{c\};\{1,2,3\}\}$. □

*Proposition 4.* The state $\rho_B$ is PPT with respect to the bipartition $\{\{2\};\{c,1,3\}\}$ if and only if $u_{B|101,110}=0$ ($u_{B|110,101}=0$).

Proof. By (12), we have

$$\overline{\rho}_{B}^{\{2\}}=\frac{1}{2}\Biggl(|0\rangle\langle 0|\otimes\phi+|1\rangle\langle 0|\otimes\sum_{\substack{p,q\in P_{3,k}\\\wedge d(p)=d(q)}}\Pr(p)u_{B|q,p}^{*}|q_1 p_2 q_3\rangle\langle p_1 q_2 p_3|$$

$$+|0\rangle\langle 1|\otimes\sum_{\substack{p,q\in P_{3,k}\\\wedge d(p)=d(q)}}\Pr(p)u_{B|q,p}|p_1 q_2 p_3\rangle\langle q_1 p_2 q_3|+|1\rangle\langle 1|\otimes\phi\Biggr). \quad (14)$$

(if) Since $U_B$ is unitary, $u_{B|101,110}=0$ implies $u_{B|110,101}=0$, vice versa. Then $|u_{B|101,101}|=1$ and $|u_{B|110,110}|=1$, that is, $U_B = u_{B|101,101}|101\rangle\langle 101|+u_{B|110,110}|110\rangle\langle 110|+u_{B|111,111}|111\rangle\langle 111|$. Thus we can obtain

$$\overline{\rho}_{B}^{\{2\}}=\frac{1}{2}\Biggl(|0\rangle\langle 0|\otimes\phi+|1\rangle\langle 0|\otimes\sum_{p\in P_{3,k}}\Pr(p)u_{B|p,p}^{*}|p\rangle\langle p|$$

$$+|0\rangle\langle 1|\otimes\sum_{p\in P_{3,k}}\Pr(p)u_{B|p,p}|p\rangle\langle p|+|1\rangle\langle 1|\otimes\phi\Biggr)$$

$$=\frac{1}{2}\left(|0\rangle\langle 0|\otimes\phi+|0\rangle\langle 1|\otimes U_{B}^{*}\phi+|1\rangle\langle 0|\otimes\phi U_{B}^{T}+|1\rangle\langle 1|\otimes\phi\right). \quad (15)$$

By (13) and (14), the characteristic equation of $\overline{\rho}_{B}^{\{2\}}$ can be written into

$$\left|\lambda I^{\otimes 4}-\overline{\rho}_{B}^{\{2\}}\right|=\lambda^{13}\left(\lambda-\frac{1}{d^2}\right)\left(\lambda-\frac{1}{d^2}\right)\left(\lambda-1+\frac{2}{d^2}\right)=0$$

Clearly, eigenvalues of $\overline{\rho}_B^{\{2\}}$ are also the same as ones of $\rho_B$. Thus the state $\rho_B$ is PPT with respect to the partition $\{\{2\};\{c,1,3\}\}$.

(only if) Assume that $u_{B|101,110} \neq 0$. Then $\langle 0100|\overline{\rho}_B^{\{2\}}|1111\rangle = \Pr(110)u_{B|101,110} \neq 0$ according to (14). It is clear that $\langle 0100|\overline{\rho}_B^{\{2\}}|0100\rangle = 0$. Thus the state $\rho_B$ is non-PPT, which is contradiction with the condition that $\rho_B$ is PPT. Thus $u_{B|101,110}=0$, which implies that $u_{B|110,101}=0$. □

Similarly, we can also obtain the following proposition.

*Proposition 5.* The state $\rho_B$ is PPT with respect to the bipartition $\{\{3\};\{c,1,2\}\}$ if and only if $u_{B|101,110}=0$ ($u_{B|110,101}=0$).

It is well known that the Peres entanglement criterion is not a characterization. This means that a PPT state may still be entangled. Thus one might ask whether the state $\rho_B$ in the re-described algorithm is entangled when it is PPT. Our answer is "no", which is shown as follows.

*Corollary 6.* For the state $\rho_B$ there is no entanglement between $\{c\}$ and $\{1,2,3\}$.

Proof. According to the proposition 3, $\rho_B$ is PPT. It is clear that $Tr_{\{1,2,3\}}(\rho_B)=\phi$ according to (12). Then it is true that $ran\left[Tr_{\{1,2,3\}}(\rho_B)\right]=ran(\phi)=3$ and $ran\left[Tr_{\{c\}}(\rho_B)\right]\leq 2$. Thus we can obtain $ran\left[Tr_{\{1,2,3\}}(\rho_B)\right]ran\left[Tr_{\{c\}}(\rho_B)\right]\leq 6$, which implies that the state $\rho_B$ is separable with respect to the bipartition $\{\{c\};\{1,2,3\}\}$. □

*Corollary 7.* The state $\rho_B$ is separable with respect to the bipartition $\{\{2\};\{c,1,3\}\}$ if and only if $u_{B|101,110}=0$ ($u_{B|110,101}=0$).

Proof. (only if) Since $\rho_B$ is separable with respect to the bipartition $\{\{2\};\{c,1,3\}\}$, it is PPT. According to the proposition 4, it is clear that $u_{B|101,110}=0$. (if) Since $u_{B|101,110}=0$, the state $\rho_B$ is PPT. Then it is true that $\rho_B$ is not distillable. We have

$$\max\left\{ran\left[Tr_{\{c,1,3\}}(\rho_B)\right], ran\left[Tr_{\{2\}}(\rho_B)\right]\right\} \leq ran(\rho_B) = ran(\rho_0) = 3,$$

which implies that $ran\left[Tr_{\{c,1,3\}}(\rho_B)\right]\leq 3$ and $ran\left[Tr_{\{2\}}(\rho_B)\right]\leq 2$. Since $ran\left[Tr_{\{c,1,3\}}(\rho_B)\right]ran\left[Tr_{\{2\}}(\rho_B)\right]\leq 6$, the $\rho_B$ is separable. □

Similarly, we can also obtain the following corollary.

*Corollary 8.* The state $\rho_B$ is separable with respect to the bipartition $\{\{3\};\{c,1,2\}\}$ if and only if $u_{B|101,110}=0$ ($u_{B|110,101}=0$).

Now we discuss the role which $\sigma_j$ and $\sigma_j^{-1}$ for $j=1,2$ played in entanglement of $\rho_B$. The above two corollaries show that $u_{B|101,110}u_{B|110,101} \neq 0$ is a sufficient and necessary conditions for entanglement of $\rho_B$ with respect to the bipartition between the second (third) work qubit and the other ones. This implies that $\sigma_1$ and $\sigma_1^{-1}$ can't change the bipartite entanglement of $\rho_B$, that is, for every bipartition $\rho_B$ is entangled if and only if $\rho_{B'}$ is entangled where $B'=\sigma_1^s B \sigma_1^t$ for any two integers $s$, $t$. In fact, it is clear that $u_{B'|101,110} = u_{B|101,110}$ and $u_{B'|110,101} = u_{B|110,101}$ according to (3), (5), and (11).

Moreover, we can also obtain that if $\rho_B$ is entangled with respect to $\{\{2\};\{c,1,3\}\}$ ($\{\{3\};\{c,2,3\}\}$) then $B$ must include $\sigma_2$ or $\sigma_2^{-1}$. The converse of this proposition is not true, which is nontrivially shown as follows.

*Proposition 9.* If $B = \sigma_2\sigma_1\sigma_2^j\sigma_1\sigma_2$ for $j \in \{0,2,4,...\}$, then $\rho_B$ is separable with respect to all bipartitions.

Proof. When $j=0$, it is clear that $B = \sigma_2\sigma_1\sigma_1\sigma_2$. According to (3), we can obtain that

$$\tau_A(B) = \tau_A(\sigma_2)\tau_A(\sigma_1)\tau_A(\sigma_1)\tau_A(\sigma_2)$$
$$= (AE_2 + A^{-1}\mathbf{1})(AE_1 + A^{-1}\mathbf{1})(AE_1 + A^{-1}\mathbf{1})(AE_2 + A^{-1}\mathbf{1})$$
$$= (A^{-2}-A^2)E_1 + (A^{-2}-A^6)E_2 + (1-A^4)E_1E_2 + (1-A^4)E_2E_1 + A^{-4}\mathbf{1}.$$

Furthermore, we have

$$U_B = (A^{-2}-A^2)\Phi(E_1) + (A^{-2}-A^6)\Phi(E_2)$$
$$+ (1-A^4)\Phi(E_1)\Phi(E_2) + (1-A^4)\Phi(E_2)\Phi(E_1) + A^{-4}I^{\otimes 3}.$$

According to (5), (6), and (11), it is clear that

$$u_{B|101,110} = (A^{-2}-A^6)\frac{\sqrt{\lambda_1\lambda_3}}{\lambda_2} + (1-A^4)\frac{\sqrt{\lambda_1\lambda_3}}{\lambda_1}$$
$$= \frac{\sqrt{\lambda_1\lambda_3}}{\lambda_2}\left[(A^{-2}-A^6) + (1-A^4)\frac{\lambda_2}{\lambda_1}\right]$$
$$= \frac{\sqrt{\lambda_1\lambda_3}}{\lambda_2}\left[(A^{-2}-A^6) + (1-A^4)d\right] = 0.$$

Assume that the state $\rho_B$ with $B = \sigma_2\sigma_1\sigma_2^j\sigma_1\sigma_2$ is separable. By (1), we can obtain

$$B' = \sigma_2\sigma_1\sigma_2^{j+2}\sigma_1\sigma_2 = \sigma_2\sigma_1\sigma_2\sigma_2^j\sigma_2\sigma_1\sigma_2 = \sigma_1\sigma_2\sigma_1\sigma_2^j\sigma_1\sigma_2\sigma_1 = \sigma_1 B \sigma_1.$$

Then the state $\rho_{B'}$ is also separable since $\rho_B$ is separable. □

Let $B = \sigma_2\sigma_1\sigma_1\sigma_2$ and $B' = \sigma_2\sigma_1\sigma_2^2\sigma_1\sigma_2$. According to (7), we can obtain that $tr[\tau_A(B)] = (A^{-8} + 2 + A^8)/d^2$ and $tr[\tau_A(B')] = (A^{-10} + A^{-2} + 2A^6)/d^2$ by simple calculation. Furthermore, it is clear that there exists an integer $k \geq 5$ such that $V_{B^{tr}}(A^{-4}) \neq V_{B'^{tr}}(A^{-4})$ by (4). Thus $B^{tr}$ and $B'^{tr}$ are topologically different while $\rho_B$ and $\rho_{B'}$ are separable for all bipartitions according to the above proposition.

One can ask whether braids whose trace closures are topologically identical have the same bipartite entanglement properties in AJL's algorithm. Our answer is yet "no". It is very interesting that there exit two braids $B, B' \in B_3$ such that $B^{tr}$ is topologically identical with $B'^{tr}$ and $\rho_B$ is separable while $\rho_{B'}$ is entangled with respect to $\{\{2\};\{c,1,3\}\}$ ($\{\{3\};\{c,1,2\}\}$). In fact, suppose that $B = \sigma_1^s$ and $B' = \sigma_2^s$ for a given positive integer $s$. It is known that $B^{tr}$ and $B'^{tr}$ are topologically identical as shown Fig. 6. By simple calculation, we can obtain $u_{B|101,110} = 0$ and $u_{B'|101,110} \neq 0$. Thus $\rho_B$ is separable for all bipartitions while $\rho_{B'}$ is entangled with respect to $\{\{2\};\{c,1,3\}\}$ ($\{\{3\};\{c,1,2\}\}$).

V. Conclusion

We firstly re-describes AJL's algorithm for three-strand braids as an algorithm which involves three work qubits in some mixed state coupled to a single control qubit. We also prove that this re-described algorithm is equivalent with AJL's algorithm. It is clear that $\rho_0$ and $\rho_1$ present in this re-described algorithm are full separable. We study the bipartite entanglement features of the state $\rho_B$ for all bipartitions by means of the Peres entanglement criterion. Our main results include: i) the state $\rho_B$ is a product state relative to the bipartition between the first work qubit and the others ; ii) there is no entanglement between the control qubit and work ones; iii) the state $\rho_B$ is entangled with respect to the bipartition between the second (third) work qubit and the others if and only if $u_{B|101,110}u_{B|110,101} \neq 0$. Moreover, we also find that braids whose trace closures are topologically identical might have different entanglement properties in AJL's algorithm.

There are several problems, however, which remain open and are worth further studies: i) in this paper we only investigate the bipartite entanglement properties in AJL's algorithm for three-strand braids. Both Multipartite entanglement properties in the algorithm for three-strand braids and entanglement ones in the algorithm for other braids are yet not clear; ii) the circuit in the re-described algorithm is same as *deterministic quantum computation with one quantum bit (DQC1)* [22] but for the initial state. In DCQ1, the initial state on work qubits is the completely mixed state. Ref. [23] has suggested that quantum discord might explain the speed-up in the DQC1. It is natural to ask whether quantum discord [24] in the re-described algorithm does the same as one in the DQC1. This problem is still open.

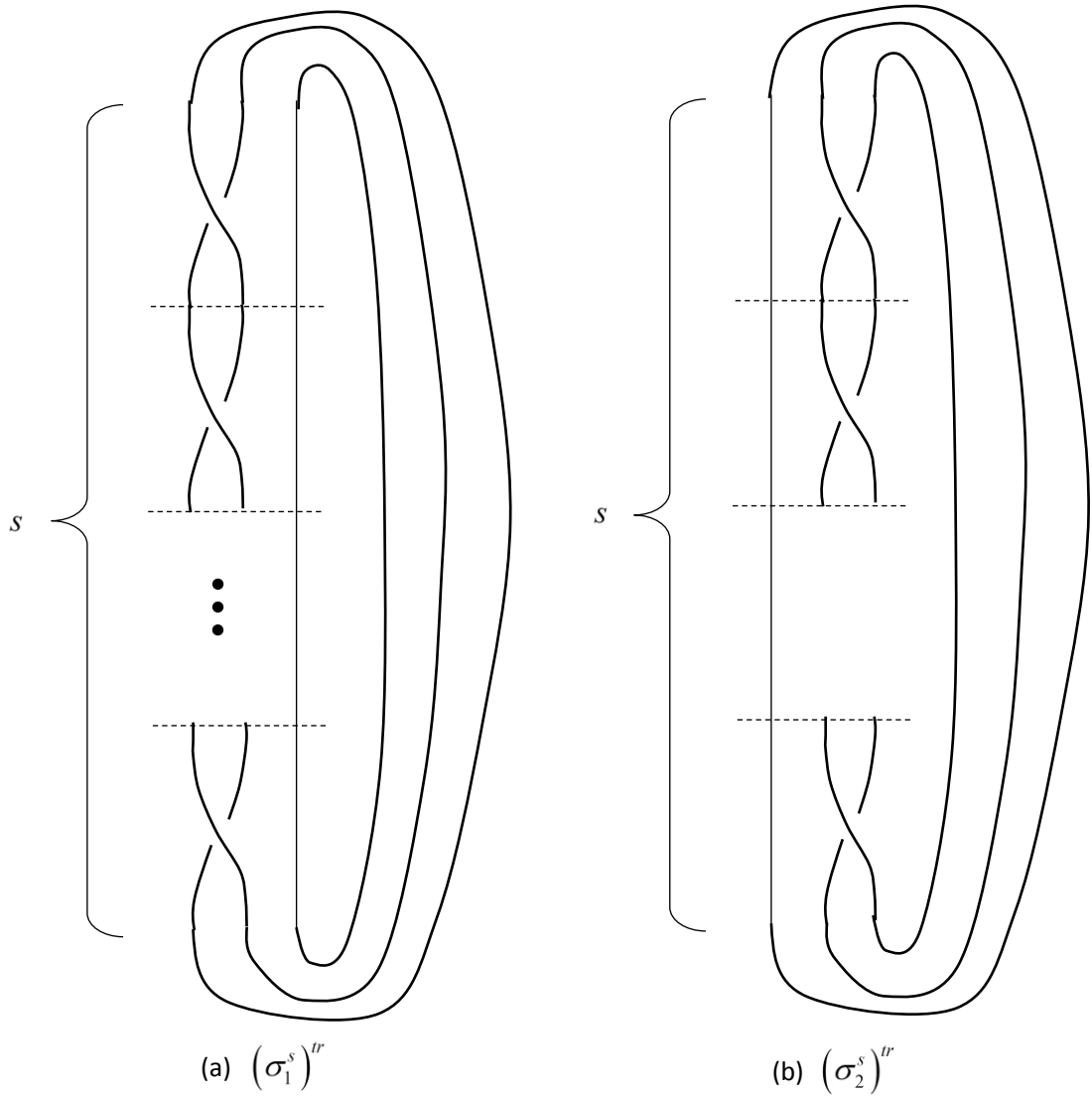

(a) $\left(\sigma_1^s\right)^{tr}$  (b) $\left(\sigma_2^s\right)^{tr}$

Fig. 6. Two three-strand braids $\sigma_1^s$, $\sigma_2^s$ and their trace closures.

**ACKNOWLEDGMENTS**

This work is supported by the Chinese National Program on Key Basic Research Project (973 Program, Grant No. 2014CB744605).